\documentclass[12pt,a4paper, fleqn]{article}

\usepackage{amsmath}
\usepackage{amssymb}
\usepackage[nosort]{cite}
\usepackage[hyperref,bulletsep]{collect}

\makeatletter
\def\mr@ignsp#1 {\ifx\:#1\@empty\else #1\expandafter\mr@ignsp\fi}%
\newcommand{\multiref}[1]{\begingroup
\xdef\mr@no@sparg{\expandafter\mr@ignsp#1 \: }%
\def\mr@comma{}%
\@for\mr@refs:=\mr@no@sparg\do{\mr@comma\def\mr@comma{,}\ref{\mr@refs}}%
\endgroup}
\makeatother

\newcommand{\hypref}[2]{\ifx\href\asklfhas #2\else\href{#1}{#2}\fi}

\renewcommand{\eqref}[1]{(\multiref{#1})}

\makeatletter
\@addtoreset{equation}{section}
\makeatother

\makeatletter
\let\old@startsection=\@startsection
\renewcommand{\@startsection}[6]{\old@startsection{#1}{#2}{#3}{#4}{#5}{#6\mathversion{bold}}}
\makeatother

\ifx\href\asklfhas\newcommand{\href}[2]{#2}\fi
\newcommand{\arxivno}[1]{\href{http://arxiv.org/abs/#1}{#1}}

\newcommand{\gym}{g_{\scriptscriptstyle\mathrm{YM}}}
\newcommand{\sfrac}[2]{{\textstyle\frac{#1}{#2}}}
\newcommand{\half}{\sfrac{1}{2}}
\newcommand{\alg}[1]{\mathfrak{#1}}
\newcommand{\cder}{\mathcal{D}}

\newcommand{\alU}{\alg{u}}
\newcommand{\alSU}{\alg{su}}

\newcommand{\alSO}{\alg{so}}
\newcommand{\alPSU}{\alg{psu}}

\newcommand{\alsl}{\alg{sl}}

\newcommand{\len}{\alg{L}}
\newcommand{\algD}{\alg{D}}
\newcommand{\algdD}{\delta\alg{D}}
\newcommand{\algP}{\alg{P}}
\newcommand{\algB}{\alg{B}}
\newcommand{\algC}{\alg{C}}
\newcommand{\algJ}{\alg{J}}
\newcommand{\algQ}{\alg{Q}}
\newcommand{\algq}{\alg{q}}

\newcommand{\algL}{\alg{L}}
\newcommand{\algS}{\alg{S}}
\newcommand{\algT}{\alg{T}}

\newcommand{\algR}{\alg{R}}
\newcommand{\algK}{\alg{K}}
\newcommand{\algj}{\alg{j}}
\newcommand{\algy}{\alg{y}}
\newcommand{\algh}{\alg{h}}
\newcommand{\algX}{\alg{X}}
\newcommand{\algx}{\alg{x}}

\newcommand{\lstate}[1]{\langle#1|}
\newcommand{\state}[1]{|#1\rangle}%

\newcommand{\comm}[2]{\left[#1,#2\right]}
\newcommand{\acomm}[2]{\left\{#1,#2\right\}}
\newcommand{\bigbrk}[1]{\bigl(#1\bigr)}

\def\[{\begin{equation}}
\def\]{\end{equation}}

\begin{document}

\setcounter{page}{0}
\thispagestyle{empty}
\begin{flushright}\footnotesize
\texttt{\arxivno{hep-th/0511109}} \\
\texttt{PUTP-2182}\\
\vspace{0.5cm}
\end{flushright}
\vspace{0.5cm}

\begin{center}
{\Large\textbf{\mathversion{bold}
$\mathcal{N}=4$ SYM to Two Loops: \\ Compact Expressions for the Non-Compact Symmetry Algebra of the $\alSU(1,1|2)$ Sector }\par}
\vspace{2cm}

\textsc{Benjamin I. Zwiebel}
\vspace{5mm}

\textit{Joseph Henry Laboratories\\
Princeton University\\
Princeton, NJ 08544, USA}
\vspace{3mm}

\texttt{bzwiebel@princeton.edu}\par\vspace*{2cm}

\textbf{Abstract}\vspace{2mm}

\begin{minipage}{12.7cm}\small
We begin a study of higher-loop corrections to the dilatation generator of $\mathcal{N}=4$ SYM in non-compact sectors. In these sectors, the dilatation generator contains infinitely many interactions, and therefore one expects very complicated higher-loop corrections. Remarkably, we find a short and simple expression for the two-loop dilatation generator. Our solution for the non-compact $\alSU(1,1|2)$ sector consists of nested commutators of four $\mathcal{O}(g^1)$ generators and one simple auxiliary generator. Moreover, the solution does not require the planar limit; we conjecture that it is valid for any gauge group. To obtain the two-loop dilatation generator, we find the complete $\mathcal{O}(g^3)$ symmetry algebra for this sector, which is also given by concise expressions. We check our solution using published results of direct field theory calculations. By applying the expression for the two-loop dilatation generator to compute selected anomalous dimensions and the bosonic $\alsl(2)$ sector internal S-matrix, we confirm recent conjectures of the higher-loop Bethe ansatz of \texttt{hep-th/0412188}.
 
\par
\end{minipage}\vspace*{\fill}

\end{center}

\newpage
\setcounter{footnote}{0}
\section{Introduction}
Our understanding of the conjectured AdS-CFT correspondence \cite{Maldacena:1997re, Gubser:1998bc, Witten:1998qj} has deepened dramatically following the observation that strings with large quantum numbers can be mapped to certain subsets of operators in the gauge theory  \cite{Berenstein:2002jq, Gubser:2002tv}. Direct progress has been made in testing the correspondence perturbatively by comparing anomalous dimensions of gauge theory operators and energies of classical and quantum corrected string solitons \cite{Frolov:2002av, Frolov:2003tu, Beisert:2003xu, Frolov:2003xy, Callan:2003xr, Beisert:2003ea, Kruczenski:2003gt, Kruczenski:2004kw, Callan:2004uv}. For reviews see \cite{Tseytlin:2003ii, Beisert:2004yq, Tseytlin:2004xa, Zarembo:2004hp, Swanson:2005wz, Plefka:2005bk}. 

These comparisons have used and motivated much progress in computing gauge theory anomalous dimensions. Beginning with the proof of one-loop integrability in the $\alSO(6)$ (scalar) sector\footnote{Integrability in four-dimensional Yang-Mills theory was first observed by Lipatov \cite{Lipatov:1994yb, Lipatov:1997vu}.} \cite{Minahan:2002ve}, computations of anomalous dimensions in planar $\mathcal{N}$ = 4 SYM have been greatly simplified by mapping single-trace gauge theory operators to states of integrable closed spin chains. This mapping allows the use of a Bethe ansatz.  In \cite{Beisert:2003yb}, integrability of the complete one-loop planar gauge theory was proven, and the corresponding Bethe ansatz was presented. After evidence of integrability was obtained via higher-loop computations of the dilatation generator in compact sectors \cite{Beisert:2003tq, Beisert:2003ys}, higher-loop Bethe ans\"{a}tze were proposed at various orders and sectors \cite{Serban:2004jf, Beisert:2004hm, Staudacher:2004tk}. This line of research culminated with a proposal for the all-loop asymptotic $\alPSU(2,2|4)$ Bethe ansatz \cite{Beisert:2005fw}. A proof of this ansatz assuming integrability was found recently \cite{Beisert:2005tm}. Parallel developments have occurred on the string side. Classical string theory in $AdS_5 \times S^5$ was shown to be integrable \cite{Mandal:2002fs,Bena:2003wd}, integrability was used to solve the classical spectrum in terms of algebraic curves \cite{Kazakov:2004qf, Beisert:2005bm}, and Bethe ans\"{a}tze for quantum strings were proposed \cite{Arutyunov:2004vx, Staudacher:2004tk, Beisert:2005fw}. 

Despite all this progress and much evidence, there is no rigorous proof of higher-loop integrability for any non-compact sector of the gauge theory. Finding the dilatation generator would be a first step towards a proof. However, even that step seems intractable for non-compact sectors. Completing a direct diagrammatic calculation is realistic only at low loop order, as was done at two loops for the fermionic $\alsl(2)$ sector in \cite{Belitsky:2005bu}. A more promising approach to higher loops is Beisert's method, which takes full advantage of superconformal symmetry. However, extending Beisert's method of computing the dilatation generator from one loop for the complete theory \cite{Beisert:2003jj, Beisert:2004ry} to three loops for the $\alSU(2|3)$ sector \cite{Beisert:2003ys}, depended on compactness. A large computer algebra computation was essential.  Because the dilatation generator in non-compact sectors is built from infinitely many interactions, a brute force computation of this kind becomes impossible. 

In this paper, we overcome this obstacle by developing techniques for higher-loop non-compact sectors. Only using constraints from Feynman diagrammatics and superconformal symmetry, we compute the two-loop dilatation operator for the non-compact $\alSU(1,1|2)$ sector. We also find the corrections to this sector's symmetry algebra up to $\mathcal{O}(g^3)$. We introduce an auxiliary generator that satisfies special commutation relations with the classical and half-loop symmetry generators. This extension of the symmetry algebra enables us to find and verify solutions of the symmetry constraints at one and one-half loops only using the commutation relations of the extended algebra at zero and at one-half loops. Via our method, higher-loop computations reduce to straightforward algebraic manipulations of commutators. This can be done efficiently even without a computer.

For our computation, it is essential that the $\alSU(1,1|2)$ sector has a hidden $\alPSU(1|1)^2$ symmetry \cite{Beisert:2004ry}, which adds tight restrictions. The representation for the $\alPSU(1|1)^2$ symmetry is trivial at leading order and has an expansion in odd powers of $g$, which is proportional to $\sqrt{\lambda}$. The generators, labeled $\overrightarrow{\algT}$ and $\overleftarrow{\algT}$,\footnote{Actually, there are four generators, $\overrightarrow{\algT}^\pm$ and $\overleftarrow{\algT}^\pm$, but the additional sign index is unimportant for the points we make here.} change the length of the spin chain, reflecting the dynamic aspect of the full $\alPSU(2,2|4)$ spin chain.  $\overrightarrow{\algT}$ and $\overleftarrow{\algT}$ anti-commute to the dilatation operator. Therefore, we only need one and one-half loops to find the two-loop dilatation operator.  More precisely, using a subscript $n$ for the $\mathcal{O}(g^n)$ corrections,  \begin{equation} \half \algdD_4= \acomm{\overrightarrow{\algT_3}}{\overleftarrow{\algT_1}}+\acomm{\overrightarrow{\algT_1}}{\overleftarrow{\algT_3}}. \end{equation}

The $\mathcal{O}(g^3)$ solution is built only from the leading (non-vanishing) order representations of the generators and the auxiliary generator, $\algh$, that acts using the harmonic numbers. Schematically, we have\footnote{The right side of the second equation actually has two terms like the one shown, using the two pairs of generators with opposite signs and directions of arrows.}  \begin{gather} \algT_3 = \pm \comm{\algT_1}{\algx}, \notag \\ \algx \sim \acomm{\overrightarrow{\algT_1}}{\comm{\overleftarrow{\algT_1}}{\algh}}. \end{gather} It follows that $\algdD_4$ is built only from the $\algT_1$'s and $\algh$.

Our solution lifts consistently and naturally to non-planar $\mathcal{N}=4$ gauge theory as well\footnote{I thank Niklas Beisert for explaining this to me.}, that is for any choice of the gauge group. In particular, it includes wrapping interactions. These are non-planar contributions to the dilatation generator that survive the planar limit by wrapping around short operators \cite{Beisert:2004hm}. Because supersymmetry relates shorter states to longer states, wrapping interactions do not contribute until four loops \cite{Beisert:2003ys, Beisert:2005fw}. However, generalizing this solution for wrapping interactions to higher loops would provide a missing piece for the comparisons between gauge and string theory. The proposed Bethe ans\"{a}tze are oblivious to corrections from wrapping interactions\footnote{For recent promising work on the nature of wrapping interactions in AdS-CFT see \cite{Ambjorn:2005wa}. Also see \cite{Sieg:2005kd}.}. 

We also find compelling evidence that integrability persists at two loops in non-compact sectors.  The two-loop dilatation generator generates the same bosonic $\alsl(2)$ subsector S-matrix as assumed for the Bethe ansatz in \cite{Staudacher:2004tk}, and the same anomalous dimensions as computed using this Bethe ansatz in \cite{Staudacher:2004tk}. A complementary approach to confirming this Bethe ansatz will appear in \cite{Eden:2005xx}.  Evidence for two-loop integrability in the fermionic $\alsl(2)$ sector was given in \cite{Belitsky:2005bu}.

Section \ref{sec:psu112} introduces the $ \alSU(1,1|2)$ sector and the residual symmetry algebra. Section \ref{sec:g2} discusses the $\mathcal{O}(g^2)$ solution, and Section \ref{sec:g3} discusses the $\mathcal{O}(g^3)$ solution and presents the two-loop dilatation generator. For simplicity, we assume planarity until Section \ref{sec:non-planar}, and in that section we present the lift to the finite-$N$ solution. In Section \ref{sec:checks}, after verifying that our solution predicts the same anomalous dimensions as those of the field theory calculations of \cite{Beisert:2003tq, Kotikov:2003fb, Eden:2005bt, Belitsky:2005bu}, we summarize the applications of the solution to compute the bosonic $\alsl(2)$ subsector S-matrix and some anomalous dimensions.  We conclude and discuss directions for further research in Section \ref{sec:outlook}. The appendix presents details about the symmetry algebra and proofs of our solution. Finally, we use many results and notations of \cite{Beisert:2004ry}.

\section{The $ \alSU(1,1|2)$ sector \label{sec:psu112}}
As explained in \cite{Beisert:2004ry}, it is consistent to restrict to various sectors of the states of $\mathcal{N}=4$ SYM. Under such a restriction, the full $\alPSU(2,2|4)$ algebra  splits into three components. One component annihilates all the states in the subsector, and the second component maps states in the subsector out of the subsector. The third component, a subalgebra, acts within the subsector non-trivially, and gives the sector its name. For the case we consider in this work, this subalgebra is $\alU(1)^2 \ltimes (\alPSU(1,1|2) \times \alPSU(1|1)^2) \ltimes \alU(1)$. The $\alU(1)^2$ consists of two external automorphisms, the length $L$ and the hypercharge $B$. The $\alU(1)$, is the quantum correction to the dilatation generator, $\algdD$, which appears as the central charge for both $\alPSU(1,1|2)$ and $\alPSU(1|1)^2$.\footnote{We label this sector with $\alSU(1,1|2)$ since this equals $\alPSU(1,1|2)  \ltimes \alU(1)$, which is the minimal algebra containing the full manifest symmetry.} Furthermore, the $\alPSU(1|1)^2$ acts trivially classically. Now we describe the restriction to this subsector and the corresponding symmetry algebra, and we present the leading non-vanishing actions of the algebra on the states of the subsector. 

\subsection{The restriction to the $ \alSU(1,1|2)$ sector} \label{sec:restriction}
To restrict to this sector, we must set the classical dimensions of states simultaneously equal to the following linear combinations of the eigenvalues of the Cartan generators of the $ \alPSU(2,2|4)$ algebra \cite{Beisert:2004ry}, 
\begin{eqnarray} \label{eq:restriction}
D_0 &=& s_1+\frac{1}{2}q_2+p+\frac{3}{2}q_1=s_2+\frac{1}{2}q_1+p+\frac{3}{2}q_2 \end{eqnarray} 
Here $[q_1, \, p, \, q_2]$ are the Dynkin labels of the $\alSU(4)$ subalgebra and $[s_1, \, s_2]$  are the Dynkin labels of the Lorentz algebra. Combined with the bounds given by the field content, this implies that $D_0$ also satisfies \begin{eqnarray} D_0 &=& L-B+s_1 = L+B+s_2.\end{eqnarray}  

\subsection{The $\alPSU(1,1|2)$ algebra \label{sec:psu112alg}}
We introduce a notation for the subset of the $\alPSU(2,2|4)$ generators that generates the $\alPSU(1,1|2)$ algebra, 
\begin{gather}\label{eq:pSU112Gen}
\algJ^0(g)=- \len+2\algD_0+\algdD(g), \quad
\algR^0=\algR^2_2-\algR^3_3, \notag \\ \algJ^{++}(g)=\algP_{22}(g),\quad
\algJ^{--}(g)=\algK^{22}(g),\quad
\algR^{\uparrow\uparrow}(g)=\algR^3_2, \quad
\algR^{\downarrow\downarrow}(g)=\algR^2_3, \notag \\
\overrightarrow{\algQ}^{+\downarrow}(g)=\algQ^{2}_2(g),\quad
\overleftarrow{\algQ}^{-\uparrow}(g)=\algS_{2}^2(g),\quad
\overrightarrow{\algQ}^{+\uparrow}(g)=\algQ^{3}_2(g),\quad
\overleftarrow{\algQ}^{-\downarrow}(g)=\algS_{3}^2(g),\notag \\
\overleftarrow{\algQ}^{+\uparrow}(g)=\dot{\algQ}_{22}(g),\quad
\overrightarrow{\algQ}^{-\downarrow}(g)=\dot{\algS}^{22}(g), \quad
\overleftarrow{\algQ}^{+\downarrow}(g)=\dot{\algQ}_{23}(g),\quad
\overrightarrow{\algQ}^{-\uparrow}(g)=\dot{\algS}^{32}(g).
\end{gather}

$\len$ is the length operator; it multiplies a state composed of $L$ fundamental fields by $L$. $\len$ commutes with all of the $\alPSU(1,1|2)$ generators. In this sector, it satisfies \begin{equation} \len =\algL^2_2-\dot{\algL}^2_2-2\algR^1_1. \end{equation} $\algL^\alpha_\beta$ and $\dot{\algL}^{\dot{\alpha}}_{\dot{\beta}}$ are Lorentz rotations. 
$\algD_0$ is the classical dilatation generator, and $\algdD$ is its quantum correction. $\algdD$'s leading term is at $\mathcal{O}(g^2)$, and its expansion includes even powers of $g$ only. The remaining generators appearing on the right side of (\ref{eq:pSU112Gen}) are the following $\alPSU(2,2|4)$ generators: $\alSU(4)$ internal R-symmetry rotation generators $\algR$, Lorentz translations and boosts $\algP$ and $\algK$, and fermionic supertranslations and superboosts $\algQ$, $\dot{\algQ}$, $\algS$, and $\dot{\algS}$. Appendix D of \cite{Beisert:2004ry} gives a complete description of the full algebra.

Note that the superscript signs correspond to $\alSU(1,1)$ charge (descended from the $\alSU(2,2)$ Lorentz subalgebra). A generator adds dimension equal to $\half$ (-$\half$) of the number of its plus (minus) signs.  Similarly, vertical arrows correspond to integer $\alSU(2)$ R-charge, and horizontal arrows correspond to half-integer hypercharge, $\algB$.
Throughout this paper we will work in a basis such that hermitian conjugation requires  switching signs and reversing arrows simultaneously. Generators without any arrows or a sign are hermitian.

The $\alPSU(1,1|2)$ algebra is given in Appendix \ref{sec:psu112comm}. However, many commutators\footnote{Note that, for simplicity, we call both commutators and anti-commutators, commutators; of course, the ``commutator'' of two fermionic generators is actually an anti-commutator.} can be inferred directly from the notation because the three types of charges are conserved. These conservation rules immediately imply that many commutators vanish. Also, $\algJ^0$ measures $\alSU(1,1)$ charge, and $\algR^0$ measures $\alSU(2)$ charge.

\subsection{The $\alPSU(1|1)^2$ algebra}
\label{sec:psu11^2alg}

The $\alPSU(1|1)^2$ algebra is generated by 
\begin{equation} \label{eq:psu11^2Gen}
\overrightarrow{\algT}^+(g)=\dot{\algQ}^{14}(g),\quad 
\overleftarrow{\algT}^-(g)=\dot{\algS}_{14}(g), \quad
\overrightarrow{\algT}^-(g)=\algS^1_1(g),\quad 
\overleftarrow{\algT}^+(g)=\algQ^1_1(g).
\end{equation}

As before, the horizontal arrows correspond to hypercharge. However, these generators carry  no $\alSU(1,1)$ charge, and the sign corresponds instead to their commutators with the external automorphism $\len,$
\begin{equation} \label{eq:length}
\comm{\len}{\algT^{\pm}(g)}=\pm \algT^{\pm}(g). \end{equation}

 The non-zero commutators are
\begin{align}\label{eq:psU11^2Alg}
\acomm{\overrightarrow{\algT}^+(g)}{\overleftarrow{\algT}^-(g)}&=\half \algdD(g), & \acomm{\overrightarrow{\algT}^-(g)}{\overleftarrow{\algT}^+(g)}&=\half \algdD(g).
\end{align}

Any commutator between non-conjugate $\algT$'s vanishes, including the squares of the $\algT$.  The product structure of the full symmetry algebra will be used many times in the rest of this work: the generators of $\alPSU(1,1|2)$ and $\alPSU(1|1)^2$ commute with each other. 

\subsection{Fields and states}
\label{sec:States}

The fields in this sector are derivatives $\cder=\cder_{22}$ acting on 
the fermions $\overrightarrow{\psi}=\Psi_{42}$ and $\overleftarrow{\psi}=\dot{\Psi}^1_2$, or the bosons $\phi^\downarrow=\Phi_{34}$ and $\phi^\uparrow=\Phi_{24}$.
We denote $k$ derivatives by a subscript $k$, 
\begin{equation} \phi^\updownarrow_k\sim\cder^k \phi^\updownarrow, \quad \overleftrightarrow{\psi_k}\sim\cder^k \overleftrightarrow{\psi}, \qquad (k \geq 0). \end{equation}

The representation of the symmetry algebra acts on a spin chain. The states of the spin chain are tensor products  \begin{equation} \state{X_1X_2 \ldots X_n} \quad \mbox{where} \quad X_i \in \left\{\phi^\updownarrow_k, \, \overleftrightarrow{\psi}_k \right\}. \end{equation}  A generic state is a linear combination of these tensor products, with the cyclic identification 
\begin{equation} \state{X_1 \ldots X_i X_{i+1} \ldots X_n}= (-1)^{(X_1\ldots X_i)(X_{i+1} \ldots X_n)} \state{X_{i+1} \ldots X_n X_1 \ldots X_i} \end{equation} $(-1)^{AB}$ is $-1$ if both $A$ and $B$ are fermionic, and $1$ otherwise.

\subsection{The leading order representation}
\label{sec:Reps}
For the leading order representation of the symmetry algebra, at $\mathcal{O}(g^0)$, the generators have one-site to one-site vertices. That is, the generators' action on the spin chain is a tensor product (given by the sum of its action on each individual site). Also, every time a fermionic generator passes a fermionic field, we must add a factor of $-1$.  The non-vanishing actions for $\algJ^{++}$, $\overrightarrow{\algQ}^{+\uparrow}$, and $\algR^{\uparrow\uparrow}$ are
\begin{align} \label{eq:FieldGen}
\algJ^{++}_0  \state{\phi^\updownarrow_k}&=(k+1)\state{\phi^\updownarrow_{k+1}},&
\algJ^{++}_0  \state{\overleftrightarrow{\psi}_k}&=\sqrt{(k+1)(k+2)}\state{\overleftrightarrow{\psi}_{k+1}},
\notag \\
\overrightarrow{\algQ_0}^{+\uparrow} \state{\phi^\downarrow_k}&=\sqrt{k+1}\state{\overrightarrow{\psi}_k},&
\overrightarrow{\algQ_0}^{+\uparrow} \state{\overleftarrow{\psi}_k}&=-\sqrt{k+1}\state{\phi^\uparrow_{k+1}}, \notag \\
\algR^{\uparrow\uparrow}\state{\phi^\downarrow_k}&=\state{\phi^\uparrow_k}.\end{align} 
The remaining parts of the leading order representation can be computed using hermitian conjugation and the algebra given in Appendix \ref{sec:psu112comm}. To find hermitian conjugates, it is simplest to use the (tensor product of the) diagonal metric  \begin{equation} \langle \phi^\downarrow_m \state{\phi^\downarrow_n} = \delta_{mn}, \quad \langle \phi^\uparrow_m \state{\phi^\uparrow_n} = \delta_{mn}, \quad \langle{\overrightarrow{\psi}_m} \state{\overrightarrow{\psi}_n} =  \delta_{mn}, \quad \langle \overleftarrow{\psi}_m \state{\overleftarrow{\psi}_n} =  \delta_{mn}. \end{equation} 
Then the condition for two operators, $\algJ$ and $\algJ^\dagger$, to be hermitian conjugates is
\begin{gather}  \lstate{X'_1 \ldots X'_m}\algJ\state{X_1\ldots X_n}= \lstate{X_1\ldots X_n}\algJ^\dagger \state{X'_1 \ldots X'_m}. \end{gather}  

\subsection{Constraints from Feynman rules}
Before beginning to discuss quantum corrections to the symmetry algebra, we review the constraints from Feynman diagrams \cite{Beisert:2004ry}. Most simply, only connected interactions appear. For the planar theory, this implies that interactions will involve replacing a set of adjacent spins of the spin chain with a new set, not necessarily of the same length. In fact, power counting implies that the $\mathcal{O}(g^n)$ term of a generator is the sum of interactions involving a total of up to $(n+2)$ initial and final spin sites.  Therefore, quantum corrections will deform the representation from the tensor product structure described in the previous subsection. The  symmetry algebra for the $\alSU(1,1|2)$ sector implies that for even n, the length of the spin chain is unchanged, and for odd n, interactions change the length of the spin chain by one. 

Feynman rules restrict to parity (or charge conjugation) even interactions. In terms of the spin chain, a parity even generator satisfies, for arbitrary fields $X_j$, 
\begin{equation} \algJ \state{X_1 \ldots X_i} = \state{X^\prime_1 \dots X^\prime_f} \Rightarrow  (-1)^{i+f_i(f_i-1)/2} \algJ \state{X_i \ldots X_1} = (-1)^{f+f_f(f_f-1)/2}\state{X^\prime_f \dots X^\prime_i} \end{equation} where $f_i$ and $f_f$ are the number of fermions in the initial and final states. 

Finally, we use the normalization for the coupling constant 
\begin{equation} g^2=\frac{\gym^2 N}{8\pi^2},\end{equation} where $N$ is the rank of the gauge group.

\subsection{The $\mathcal{O}(g^1)$ solution}
Applying the rules from the previous subsection, at $\mathcal{O}(g^1)$, the corrections will entail replacing one spin site with two, or vice-versa. For instance, a one-site to two-site generator $\algT$ would act on a generic state as:
\begin{gather} \algT \state{X_1 \ldots X_i \ldots X_n} = \notag \\ 
\sum_{a, \, b} c^{X_1}_{ab} \state {Y_a Y_b X_2 \ldots X_i \ldots X_n} + \cdots + \notag \\
(-1)^{(X_1\ldots X_{i-1})\algT} \sum_{a,\,b} c^{X_i}_{ab} \state{X_1 \ldots X_{i-1} Y_a Y_b X_{i+1} \ldots X_n} + \cdots, \end{gather}  where $Y_a$ and $Y_b$ run over all fields. For fixed $i$, only a finite number of the $c^{X_i}_{ab}$ will be non-zero.

Starting at this order, it is possible to construct interactions that vanish on closed spin chain states because of the cyclic identification. These correspond to gauge transformations in gauge theory. An example of a one-site to two-site gauge transformation satisfies, for all fields $X_i$ and a fixed field $Y$, \begin{equation} \algT \state{X_i} = \state{X_i Y} - (-1)^{Y X_i} \state{Y X_i}. \end{equation} In fact, the symmetry algebra can be satisfied at this order only up to gauge transformations \cite{Beisert:2004ry}.  

Only the $\alPSU(1|1)^2$ generators receive $\mathcal{O}(g)$ corrections since only they change the length of the spin chain. The action of $\overrightarrow{\algT}_1^{+}$ on states composed of the $\overrightarrow{\psi}_k$ was derived in \cite{Beisert:2004ry}, and it is straightforward  to generalize to the full solution by requiring commutation with the $\alPSU(1,1|2)$ algebra (up to gauge transformations).
 
\begin{eqnarray} \overrightarrow{\algT_1}^{+}\state{\phi^\updownarrow_m} &=& \frac{1}{\sqrt{2}}\sum_{k=0}^{m-1} \frac{1}{\sqrt{k+1}}\left(\state{\overrightarrow{\psi_k}\phi^\updownarrow_{m-1-k}}-\state{\phi^\updownarrow_{m-1-k}\overrightarrow{\psi_k}}\right), \notag \\  
\overrightarrow{\algT_1}^{+} \state{\overrightarrow{\psi_m}} &=& \frac{1}{\sqrt{2}}\sum_{k=0}^{m-1} \sqrt{\frac{m+1}{(k+1)(m-k)}} \state{\overrightarrow{\psi_k}\overrightarrow{\psi_m}_{-1-k}}, \notag \\ 
\overrightarrow{\algT_1}^{+}\state{\overleftarrow{\psi_m}} &=& \frac{1}{\sqrt{2}}\sum_{k=0}^{m-1} \sqrt{\frac{m-k}{(k+1)(m+1)}}\left(\state{\overrightarrow{\psi_k}\overleftarrow{\psi_m}_{-1-k}}+\state{\overleftarrow{\psi_m}_{-1-k}\overrightarrow{\psi_k}}\right) \notag \\
&-& \frac{1}{\sqrt{2(m+1)}}\sum_{k=0}^{m} \left(\state{\phi^\downarrow_k\phi^\uparrow_{m-k}}-\state{\phi^\uparrow_k\phi^\downarrow_{m-k}}\right). \end{eqnarray} 

Up to a single minus sign in the last line, $\overleftarrow{\algT_1}^{+}$ follows from switching $\overrightarrow{\psi_k}$ and $\overleftarrow{\psi_k}$ in the above expression for $\overrightarrow{\algT_1}^{+}$, 
 \begin{eqnarray} \overleftarrow{\algT_1}^{+}\state{\phi^\updownarrow_m} &=& \frac{1}{\sqrt{2}}\sum_{k=0}^{m-1} \frac{1}{\sqrt{k+1}}\left(\state{\overleftarrow{\psi_k}\phi^\updownarrow_{m-1-k}}-\state{\phi^\updownarrow_{m-1-k}\overleftarrow{\psi_k}}\right), \notag \\  
\overleftarrow{\algT_1}^{+} \state{\overleftarrow{\psi_m}} &=& \frac{1}{\sqrt{2}}\sum_{k=0}^{m-1} \sqrt{\frac{m+1}{(k+1)(m-k)}} \state{\overleftarrow{\psi_k}\overleftarrow{\psi_m}_{-1-k}}, \notag \\ 
\overleftarrow{\algT_1}^{+}\state{\overrightarrow{\psi_m}} &=& \frac{1}{\sqrt{2}}\sum_{k=0}^{m-1} \sqrt{\frac{m-k}{(k+1)(m+1)}}\left(\state{\overleftarrow{\psi_k}\overrightarrow{\psi_m}_{-1-k}}+\state{\overrightarrow{\psi_m}_{-1-k}\overleftarrow{\psi_k}}\right) \notag \\
&+& \frac{1}{\sqrt{2(m+1)}}\sum_{k=0}^{m} \left(\state{\phi^\downarrow_k\phi^\uparrow_{m-k}}-\state{\phi^\uparrow_k\phi^\downarrow_{m-k}}\right). \label{eq:leftT1+} \end{eqnarray} 
The other two $\alPSU(1|1)^2$ generators, the $\algT^-$, can then be computed at this order via hermitian conjugation. Up to gauge transformations, and a rescaling of $g$, this solution is completely fixed by the symmetry constraints.

\section{Order $g^2$ \label{sec:g2}}
At $\mathcal{O}(g^2)$, $\algdD$ and the $\alPSU(1,1|2)$ generators receive quantum corrections. From (\ref{eq:psU11^2Alg}), it is straightforward to compute \begin{equation} \algdD_2= 2 \acomm{\overrightarrow{\algT_1}^+}{\overleftarrow{\algT_1}^-}=2 \acomm{\overrightarrow{\algT_1}^-}{\overleftarrow{\algT_1}^+}.\end{equation} As first shown in \cite{Beisert:2003jj}, $\algdD_2$ acts by projecting two adjacent sites onto modules of definite $\alPSU(2,2|4)$ ``spin'' $j$ with coefficient $h(j)$. $h$ gives the harmonic numbers \begin{equation} h(k) = \sum_{k'=1}^{k}\frac{1}{k'}=\psi(k+1)-\psi(1).  \end{equation} The harmonic numbers will play an essential role in $\algdD_4$ as well. 

$\algdD$ is the only generator we need to compute because, as we now explain, once we know $\algdD$ the full $\alPSU(1,1|2) \times \alPSU(1|1)^2$ algebra's action is fixed by group theory. Knowing $\algdD$ means knowing its eigenstates and eigenvalues. Multiplets are then formed by states of equal eigenvalues, and the generators of $\alPSU(1,1|2) \times \alPSU(1|1)^2$ must connect the states of a multiplet with factors determined by group theory. 

However, for our method of computing $\algdD$, it is essential to compute the perturbative corrections to the $\alPSU(1,1|2)$ generators. They are needed for constraining the one and one-half loop $\alPSU(1|1)^2$ generators, which anti-commute to the two-loop dilatation generator. Moreover, the solution we present below for the $\alPSU(1,1|2)$ generators has an interesting and simple structure. Next we present this solution and discuss its possible modifications and its proof.

\subsection{The solution \label{sec:g2solution}}
We define two auxiliary generators that play central parts in our solution. $\algh$ is a one-site generator of harmonic numbers. Its action is 
\begin{align} \algh \state{\phi^\updownarrow_k}&=\half h(k)\state{\phi^\updownarrow_k},&  \algh \state{\overleftrightarrow{\psi_k}}&= \half h(k+1)\state{\overleftrightarrow{\psi_k}}. \end{align} $\algx$ is a two-site to two-site generator that we can write in two equivalent ways,
\begin{eqnarray} \label{eq:definex} \algx &=& \acomm{\overleftarrow{\algT_1}^-}{\comm{\overrightarrow{\algT_1}^+}{\algh}}-\acomm{\overleftarrow{\algT_1}^+}{\comm{\overrightarrow{\algT_1}^-}{\algh}} \nonumber \\
&=& \acomm{\overrightarrow{\algT}^-_1}{\comm{\overleftarrow{\algT}^+_1}{\algh}}-\acomm{\overrightarrow{\algT}^+_1}{\comm{\overleftarrow{\algT}^-_1}{\algh}}. \end{eqnarray} 
The equality in (\ref{eq:definex}) follows from (\ref{eq:psU11^2Alg}), since
\begin{equation} \acomm{\overleftarrow{\algT_1}^-}{\comm{\overrightarrow{\algT_1}^+}{\algh}}+\acomm{\overrightarrow{\algT_1}^+}{\comm{\overleftarrow{\algT_1}^-}{\algh}} = \half \comm{\algdD_2}{\algh}, \end{equation} and the analogous equation for $\overrightarrow{\algT_1}^-$ and its conjugate is satisfied. Because of this equality in (\ref{eq:definex}), $\algx$ is hermitian.

Let $\algX^\pm$ represent $\algJ^{++}$, $\algJ^{--}$, or the eight $\algQ$'s, where we retain only the $\alSU(1,1)$ charge. Then the solution of the symmetry and Feynman diagram constraints is
\begin{equation} \label{eq:g2solution} \algX^\pm_2=\pm \comm{\algX^\pm_0}{\algx} +\comm{\algX^\pm_0}{\algy}. \end{equation}
We present an outline of the proof that this satisfies the algebra relations in section \ref{sec:g2discussion} and more details in Appendix \ref{sec:g2proof}.

$\algy$ is a two-site to two-site generator that commutes with $\algB$, $\algD_0$ and the $\algR$'s. Commuting all the generators of the $\alSU(1,1|2) \times \alPSU(1|1)^2$ algebras with a generator such as $\algy$ maps one solution of the commutation relations to another. It corresponds to the first term in the expansion of the similarity transformation  \begin{equation} \algJ \mapsto U \algJ U^{-1}, \quad U = 1 + g^2 \algy + \cdots, \quad \mbox{i.e.} \quad \algJ_2 \mapsto \algJ_2 + \comm{\algJ_0}{\algy}. \end{equation} We require $\algy$ to commute with the $\algR$'s and with $\algB$ and $\algD_0$ to preserve $\algdD$'s manifest R-symmetry and $\algdD$'s eigenstates' hypercharge and classical dimension assignments. To maintain manifest consistency with the Feynman diagram rules, $U$'s expansion must be in even powers of $n$, consisting of $(\frac{n}{2}+1)$-site to $(\frac{n}{2}+1)$-site interactions. For anti-hermitian (or vanishing) $\algy$, $\algX^+$and $\algX^-$ are hermitian conjugates up to $\mathcal{O}(g^2)$.

\subsection{Freedom for the $\mathcal{O}(g^2)$ solution}
There are two possible sources of freedom for the solution at this order: interactions that vanish on cyclic states (gauge transformations) and homogeneous solutions. We now exclude the former and discuss the latter.

The requirement of even parity rules out the possibility of applying gauge transformations to the solution at this order, since generators are sums of two-site to two-site interactions. This also implies that the algebra is satisfied exactly (not just modulo gauge transformations). 

However, at this point, we cannot rule out modification by a homogeneous solution. Under this modification, \begin{equation} \algJ^{++}_2 \mapsto \algJ^{++}_2 + \delta\algJ^{++}_2, \quad \overleftrightarrow{\algQ_2}^{+\updownarrow} \mapsto \overleftrightarrow{\algQ_2}^{+\updownarrow} + \delta \overleftrightarrow{\algQ_2}^{+\updownarrow}, \end{equation} and similarly for the hermitian conjugates. In order for the symmetry constraints to remain satisfied, the $\delta \algJ$'s and $\delta \algQ$'s must not contribute to any commutator of the algebra. For example, \begin{equation} \comm{\delta\algJ^{++}_2}{\algJ^{--}_0}+ \comm{\algJ^{++}_0}{\delta\algJ^{--}_2} = 0. \end{equation} We have not found any non-trivial homogeneous solutions, or ruled them out. However, from the above discussion regarding $\algdD$, we conclude that once $\algdD_4$ is found, this freedom is fixed. We will find the solution for $\algdD_4$ below. Since the $\mathcal{O}(g^2)$ solution presented in this section is consistent with it, this is the field theory $\mathcal{O}(g^2)$ solution.

\subsection{Discussion of the proof of the $\mathcal{O}(g^2)$ solution \label{sec:g2discussion}}
To prove the solution we must check that the commutators given in Appendix \ref{sec:psu112comm} are satisfied at $\mathcal{O}(g^2)$. The proof is based upon substitution of the solution, elementary algebra, and keeping track of powers of $g$. The commutators up to $\mathcal{O}(g^1)$ and the identities presented in Appendix \ref{sec:g2properties} are also needed. In particular, many of the commutators are simplified since the entire solution is hermitian and given by commutators of leading order generators and $\algx$. In Appendix \ref{sec:g2proof} we verify a representative set of commutators. Here we do one example in detail.

The commutator of the conjugate $\algJ$'s  \begin{equation} \comm{\algJ^{--}}{\algJ^{++}}=\algJ^0 \end{equation} has the $\mathcal{O}(g^2)$ component \begin{equation}  \comm{\algJ^{--}}{\algJ^{++}}_2=\comm{\algJ^{--}_2}{\algJ^{++}_0}+ \comm{\algJ^{--}_0}{\algJ^{++}_2}=\algJ^0_2=\algdD_2.\end{equation}   Expanding the solution for $\algJ^{++}_2$ yields\footnote{We set $\algy$ to zero without loss of generality.} 
\begin{eqnarray} \algJ^{++}_2&=&\comm{\algJ^{++}_0}{\algx} \nonumber \\
&=&  \acomm{\overleftarrow{\algT_1}^-}{\comm{\overrightarrow{\algT_1}^+}{\algj^{++}}}-\acomm{\overleftarrow{\algT_1}^+}{\comm{\overrightarrow{\algT_1}^-}{\algj^{++}}},
\end{eqnarray}
where we have defined \begin{equation} \algj^{++}=\comm{\algJ_0^{++}}{\algh}. \end{equation} 
Also, after defining \begin{equation} \algj^{--}=-\comm{\algJ_0^{--}}{\algh}, \end{equation}  direct computation shows that \begin{equation} \comm{\algJ_0^{--}}{\algj^{++}}=-\half \len. \end{equation} Using these identities we find, 
\begin{eqnarray} \label{eq:J-J+} \comm{\algJ^{--}_0}{\algJ^{++}_2}&=&
\acomm{\overleftarrow{\algT}^-_1}{\comm{\overrightarrow{\algT_1}^+}{\comm{\algJ^{--}_0}{\algj^{++}}}}-\acomm{\overleftarrow{\algT_1}^+}{\comm{\overrightarrow{\algT_1}^-}{\comm{\algJ^{--}_0}{\algj^{++}}}} \nonumber \\
&=& - \half \acomm{\overleftarrow{\algT_1}^-}{\comm{\overrightarrow{\algT_1}^+}{\len}}+ \half \acomm{\overleftarrow{\algT_1}^+}{\comm{\overrightarrow{\algT_1}^-}{\len}} \nonumber \\
&=& \half \acomm{\overleftarrow{\algT_1}^-}{\overrightarrow{\algT_1}^+}+\half \acomm{\overleftarrow{\algT_1}^+}{\overrightarrow{\algT_1}^-} \nonumber \\
&=& \half \algdD_2 \end{eqnarray}
To reach the second to last line, we used (\ref{eq:length}), and for the last line (\ref{eq:psU11^2Alg}) was needed. 
By hermiticity, \begin{equation} \comm{\algJ^{--}_2}{\algJ^{++}_0}= \comm{\algJ^{--}_0}{\algJ^{++}_2}^\dagger. \end{equation} 
So, finally we have, \begin{eqnarray} \comm{\algJ^{--}}{\algJ^{++}}_2&=& \comm{\algJ^{--}_0}{\algJ^{++}_2}+\comm{\algJ^{--}_2}{\algJ^{++}_0} \nonumber \\
&=& \comm{\algJ^{--}_0}{\algJ^{++}_2}+\comm{\algJ^{--}_0}{\algJ^{++}_2}^\dagger \nonumber \\
&=& \half \algdD_2 + \half \algdD_2^\dagger \nonumber \\
&=& \algdD, \end{eqnarray}
as required by the algebra. We used the hermiticity of $\algdD_2$ to reach the last line.
\section{Order $g^3$} \label{sec:g3}
With the $\mathcal{O}(g^2)$ solution, we can use the constraints to find the $\mathcal{O}(g^3)$ solution, which consists of corrections to the $\alPSU(1|1)^2$ generators. We now present and discuss this $\mathcal{O}(g^3)$ solution and its proof, the two-loop dilatation generator that follows, and the lift to the finite-$N$ dilatation generator.
\subsection{The solution \label{sec:g3solution}}
Only the $\alPSU(1|1)^2$ generators, the $\algT$, receive corrections at this order. Once again the form of the solution depends only on the sign of the generator (though recall that now the sign refers to the commutator with $\len$). 
\begin{equation} \algT^\pm_3=\pm \comm{\algT^\pm_1}{\algx} + \comm{\algT^\pm_1}{\algy} + \alpha \algT^\pm_1.  \label{eq:g3solution} \end{equation} Again, the $\algy$ commutator is a similarity transformation, and it must be the same as that of the $\mathcal{O}(g^2)$ solution. $\alpha$ corresponds to the coupling constant transformation \begin{equation} g \mapsto g + \alpha g^3. \end{equation} As at $\mathcal{O}(g^2)$, the solution is hermitian, provided $\algy$ is anti-hermitian. It is difficult to imagine a simpler solution. Beside the coupling constant transformation and the similarity transformation, the solution at this order is just a commutator with $\algx$, as was the case for $\mathcal{O}(g^2)$.

As at the previous order, we use a direct method to prove that this solution satisfies the symmetry algebra constraints, and the proof is in Appendix \ref{sec:g3proof}. 

\subsection{Freedom for the $\mathcal{O}(g^3)$ solution}
At this order, we could add gauge transformations to the generators. Furthermore, the solution satisfies the commutation relations only up to gauge transformations.

The case for homogeneous solutions at this order exactly parallels that of the previous order.  Under a homogeneous modification, \begin{equation} \overleftrightarrow{\algT_3}^{\pm} \mapsto \overleftrightarrow{\algT_3}^{\pm} + \delta\overleftrightarrow{\algT_3}^{\pm}. \end{equation}  In order for the symmetry constraints to remain satisfied, the $\delta \algT$'s must not contribute to any commutator of the algebra, both for commutators among the $\alPSU(1|1)^2$ generators and for those with $\alPSU(1,1|2)$ generators.  Again, we have not found any non-trivial homogeneous solutions, or ruled them out. However, as for the $\mathcal{O}(g^2)$ solution, the successful checks of our solution with field theory computations implies that such a homogeneous contribution is not part of the field theory solution. 

\subsection{The two-loop dilatation operator}
From (\ref{eq:psU11^2Alg}) we can now compute $\algdD_4$ directly, \begin{equation}
\algdD_4=2 \acomm{\overrightarrow{\algT}^+}{\overleftarrow{\algT}^-}_4=2 \acomm{\overleftarrow{\algT}^+}{\overrightarrow{\algT}^-}_4. \end{equation}
It follows that $\algdD_4$ is composed only of the $\algT_1$'s and $\algh$, the one-site  harmonic number generator. After setting the similarity transformation $\algy$ to zero, without loss of generality, and using the vanishing of the squares of the $\algT$'s, we find \begin{eqnarray} \algdD_4&=& 2 \acomm{\overrightarrow{\algT_1}^+}{\comm{\overleftarrow{\algT_1}^-}{\acomm{\overleftarrow{\algT_1}^+}{\comm{\overrightarrow{\algT_1}^-}{\algh}}}}+ 2\acomm{\overleftarrow{\algT_1}^-}{\comm{\overrightarrow{\algT_1}^+}{\acomm{\overrightarrow{\algT_1}^-}{\comm{\overleftarrow{\algT_1}^+}{\algh}}}} \nonumber \\ &=&   2 \acomm{\overleftarrow{\algT_1}^+}{\comm{\overrightarrow{\algT_1}^-}{\acomm{\overrightarrow{\algT_1}^+}{\comm{\overleftarrow{\algT_1}^-}{\algh}}}}+ 2\acomm{\overrightarrow{\algT_1}^-}{\comm{\overleftarrow{\algT_1}^+}{\acomm{\overleftarrow{\algT_1}^-}{\comm{\overrightarrow{\algT_1}^+}{\algh}}}}. \nonumber \\ & &  \label{eq:D4solution} \end{eqnarray}
In this expression we have left out the coupling constant transformation parameterized by $\alpha$ in (\ref{eq:g3solution}), which leads to \begin{equation} \algdD_4 \mapsto \algdD_4 + 2 \alpha \, \algdD_2. \end{equation} However, to match field theory results $\alpha$ must be zero.
\subsection{Non-planarity and wrapping interactions \label{sec:non-planar}}
By lifting our expressions for the building blocks of $\algdD_4$ to their non-planar generalization, we will construct a candidate for the finite-$N$ $\algdD_4$. To support our conjecture that this is the correct solution, we will observe that it accurately includes wrapping interactions. The two-loop non-planar solution for the $\alSU(2)$ sector (a subsector of the $\alSU(1,1|2)$ sector) was found in \cite{Beisert:2003tq}. In that case, there is a unique lift from the planar to the non-planar theory.

The non-planar action for the one-site generators, including the $\mathcal{O}(g^0)$ terms and $\algh$ are straightforward to obtain. Let the gauge group of the theory have generators $\mathfrak{t_m}$ and metric $\mathfrak{g^{mn}}$. Then, for instance, using the notation of \cite{Beisert:2004ry} \begin{equation} \algh= \sum_{k=0}^{\infty} h(k)\, \mbox{Tr}\, (\phi^\downarrow_k \,  \check{\phi}^\downarrow_k+\phi^\uparrow_k \,  \check{\phi}^\uparrow_k)+h(k+1) \, \mbox{Tr}\, (\overleftarrow{\psi_k} \, \check{\overleftarrow{\psi_k}} + \overrightarrow{\psi_k} \, \check{\overrightarrow{\psi_k}}),\end{equation} where for $X_i \in \{\phi_k^\updownarrow,\,\overleftrightarrow{\psi_k}\}$, we have the expansion $X_i=X^\mathfrak{m}_i\mathfrak{t_m}$, and \begin{equation} \check{X_i} = \mathfrak{t}_\mathfrak{m}\mathfrak{g^{mn}}\frac{\delta}{\delta X^\mathfrak{n}_i}, \quad \frac{\delta}{\delta X^\mathfrak{m}_i}X^\mathfrak{n}_j=\delta_{ij}\delta^\mathfrak{n}_\mathfrak{m}. \end{equation} 

The $\algT_1$ also have a natural generalization for the non-planar theory. \begin{eqnarray} \overrightarrow{\algT_1}^+=&\sum_{\substack{0 \leq m \\ 0 \leq k < m}}& \frac{1}{\sqrt{2(k+1))}}\,\mbox{Tr}\,\comm{\overrightarrow{\psi_k}}{\phi_{m-1-k}^\updownarrow} \check{\phi}_m^\updownarrow \nonumber \\ + & \sum_{\substack{0 \leq m \\ 0 \leq k < m}} & \frac{m+1}{2 \sqrt{2(k+1)(m-k)}}\,\mbox{Tr}\,\acomm{\overrightarrow{\psi_k}}{\overrightarrow{\psi_m}_{-1-k}} \check{\overrightarrow{\psi_m}} \nonumber \\  + & \sum_{\substack{0 \leq m  \\ 0 \leq k < m}} & \frac{m-k}{\sqrt{2(k+1)(m+1)}}\,\mbox{Tr}\,\acomm{\overrightarrow{\psi_k}}{\overleftarrow{\psi_m}_{-1-k}} \check{\overleftarrow{\psi_m}} \nonumber \\ + & \sum_{\substack{0 \leq m \\ 0 \leq k \leq m}} & \frac{1}{\sqrt{2(m+1)}}\,\mbox{Tr}\,\comm{\phi_k^\downarrow}{\phi_{m-k}^\uparrow} \check{\overleftarrow{\psi_m}}. \end{eqnarray}

There is a similar expression for $\overleftarrow{\algT_1}^+$, which can be read from (\ref{eq:leftT1+}). For the hermitian conjugates, the $\algT_1^-$, simply perform the switch \begin{equation} X_i \leftrightarrow \check{X}_i, \quad \forall X_i. \end{equation} Substituting these expressions into the expressions for $\algx$, gives its non-planar version. Then the expressions given for $\algT_3$ and $\algdD_4$ become non-planar. Because the proof that the planar solution satisfies the symmetry constraints is independent of planarity, the non-planar generalization still satisfies the symmetry constraints.

While we do not have a proof that this is the correct non-planar solution, our solution  accurately includes wrapping interactions, which can be thought of as special cases of non-planar interactions. Wrapping interactions apply to two-site states, for which the planar solution and the non-planar generalization are equivalent. Since the $\algT_1$'s map one site to two sites or vice-versa, the action of $\algdD_4$ is well defined even on two-site states. Since acting with the $\algT^+$'s on two-site states yields three-site states, two-site states are in the same multiplets as three-site states. Therefore, adding special wrapping interactions that only change the anomalous dimensions of two-site states would be inconsistent with the symmetry constraints.  

\section{Tests and applications of the solution \label{sec:checks}}
Using the solution for the two-loop dilatation operator, we first provide strong evidence that it is correct via direct diagonalization and comparison to rigorous field theory computations. We then use our solution to present strong evidence in favor of integrability by computing the internal S-matrix in the bosonic $\alsl(2)$ sector and by comparing anomalous dimension predictions of the Bethe ansatz of \cite{Staudacher:2004tk} with the results of direct diagonalization.

\subsection{Two-loop planar anomalous dimensions}
Expanding the expression for $\algdD_4$ in terms of interactions, we find the planar anomalous dimensions by direct diagonalization.  We first identify the spin chain states of the subspaces of certain (small) values of classical dimension, R-charge, length, and hypercharge. Then we apply $g^2 \algdD_2 + g^4\algdD_4$ to these subspaces and compute its eigenvalues (the anomalous dimensions) and eigenstates. Again, we have used \texttt{Mathematica}. We check states with rigorously known anomalous dimensions. These include twist-two operators \cite{Kotikov:2003fb}, a pair of states of length three and bare dimension six \cite{Eden:2005bt}, two excitation states (BMN operators) \cite{Beisert:2002tn, Beisert:2003tq}, and length-three states built from one type of fermion and from derivatives (in the fermionic $\alsl(2)$ subsector) \cite{Belitsky:2005bu}. The states we check, given in Table \ref{tab:comparefield}, are in complete agreement with these previous computations. Therefore, we conclude that we have found the correct solution for $\algdD_4$. Since our comparison includes length-two states, we find confirmation that no additional wrapping terms are needed. \begin{table}\centering
$\begin{array}{|l|l|l|}\hline
\algD_0&(R, L, B) &\bigbrk{\algdD_2,\algdD_4}^P\\\hline
4  &(2,2,0)& \begin{array}{l} (6,-12)^+ \end{array} \\
\hline 
5  &(3,3,0)& \begin{array}{l} (4,-6)^- \end{array} \\
\hline
6  &(2,2,0)& \begin{array}{l} (\frac{25}{3},-\frac{925}{54})^+ \end{array} \\
\hline
6 & (3, 3, 0)& \begin{array}{l} (\frac{15}{2}, -\frac{225}{16})^\pm \end{array} \\ \hline
6  &(4,4,0)&\begin{array}{l} (2.76393,-2.90983)^+ \\ (7.23607,-14.0902)^+ \end{array} \\ \hline
7 & (5, 5, 0)& \begin{array}{l} (2, -\frac{3}{2})^- \\ (6, -\frac{21}{2})^- \end{array} \\ \hline 
7.5 & (0, 3, \pm \frac{3}{2})& \begin{array}{l} (10, -\frac{245}{12})^\pm \end{array} \\ \hline 
8  &(2,2,0)& \begin{array}{l} (\frac{49}{5},-\frac{45619}{2250})^+ \end{array} \\
\hline 
8 & (6, 6, 0)& \begin{array}{l} (1.50604, -0.830063)^+ \\ (4.89008, -7.30622)^+ \\ (7.60388, -14.8637)^+ \end{array} \\ \hline
9  &(6,8,\pm 1)& \begin{array}{l} (1.17157,-0.4895952)^- \\ (4, -5)^- \\ (6.82843, -12.5104)^- \end{array} \\
\hline 
9.5 & (0, 3, \pm \frac{3}{2})& \begin{array}{l} (\frac{133}{12},-\frac{131117}{5760})^\pm \end{array} \\ \hline 
10  &(7,9,\pm 1)& \begin{array}{l} (0.935822,-0.304865)^+ \\ (3.30540, - 3.44381)^+ \\ (6, -10)^+ \\  (7.75877, -15.2513)^+ \end{array} \\ \hline 
10.5 & (0, 3, \pm \frac{3}{2})& \begin{array}{l} (\frac{761}{70}, -\frac{138989861}{6174000})^+ \\ (\frac{ 761}{60},-\frac{419501}{16000})^\pm \end{array} \\ \hline \end{array}$ \caption{Two-loop spectrum for states with rigorously known planar anomalous dimensions. The $P$ exponent of the anomalous dimensions gives the states' eigenvalues under parity. The $\pm$ pairs for $P$ are a consequence of integrability. The $\pm$ pairs for $B$ come from switching the two types of fermions.  The twist-two operators are those with length two, the two excitation states satisfy $\algD_0 - L \leq 2$, and the three-fermion states have $R= 0$.}
\label{tab:comparefield}
\end{table}
\subsection{The two-loop $\alsl(2)$ S-matrix and diffractionless scattering}
We now perform a new two-loop check of the bosonic $\alsl(2)$ sector Bethe ansatz of \cite{Staudacher:2004tk}. Instead of only checking anomalous dimension predictions, we also verify a key part of its derivation, the S-matrix. 
It is straightforward to restrict to the two-excitation $\alsl(2)$ sector, consisting of states composed only of $\phi^\downarrow$'s and two or fewer derivatives\footnote{Of course, by $R$-symmetry, the sector with $\phi^\uparrow$'s has the same S-matrix and anomalous dimensions.}. We have computed the internal S-matrix as in \cite{Staudacher:2004tk}, which used ideas introduced in \cite{Bethe:1931hc} and \cite{Sutherland:1978xx}.  A basis for two excitation states is \begin{equation} 
\state{\Psi_{x_1x_2}}= \state{\ldots\phi^\downarrow\stackrel{\stackrel{x_1}{\downarrow}}{\cder(\phi^\downarrow)}\phi^\downarrow\ldots\phi^\downarrow\stackrel{\stackrel{x_2}{\downarrow}}{\cder(\phi^\downarrow)}\phi^\downarrow\ldots},  \end{equation} where the derivatives appear at sites $x_1$ and $x_2$ of the spin chain. Then, the Schr\"odinger equation \begin{equation} H \state{\Psi} = E \state{\Psi}, \qquad H = g^2 \, \algdD_2+g^4 \, \algdD_4, \end{equation} is solved by the ansatz \begin{gather} \state{\Psi} = \sum_{1 \leq x_1 \leq x_2 \leq L} \left(f_-(\delta x, p_i)e^{ip_1x_1+ip_2x_2}+f_+(\delta x, p_i)e^{ip_2x_1+ip_1x_2} \right) \state{\Psi_{x_1x_2}}, \nonumber \\  \delta x = x_2-x_1. \end{gather} $L$ is the length of the spin chain, and the $p_i$ are the momenta of the excitations which scatter off each other. Since the Hamiltonian is short-ranged and translationally invariant, for large $\delta x$ the solutions of the Schr\"odinger equation reduce to superpositions of one excitation eigenstates, proportional to $e^{i p x}$. The S-matrix gives the phase that one excitation's wave function acquires when passing the other excitation, \begin{equation} S(p_2,p_1) = \frac{f_+(\delta x, p_i)}{f_-(\delta x, p_i)}, \quad \delta x > 1. \end{equation}  The inequality reflects that the Hamiltonian has interactions involving at most three adjacent sites. This short-range Hamiltonian also leads to the following ansatz\footnote{For simplicity we drop the $p_i$, but all functions still depend on them. Note that $f_-$ and $f_0$ are unphysical. They will transform non-trivially under a similarity transformation for $\algdD$, unlike $S$ and $E$.}, \begin{equation} f_-(\delta x > 1) =  1,\, f_+(\delta x > 0) = S,\, f_-(1)=f_-, \, f_\pm(0)=f_0. \end{equation} Using $\texttt{Mathematica}$, we have solved the Schr\"odinger equation using this ansatz and our expression for the dilatation generator. The solution for the energy and the S-matrix is  \begin{gather} E= E(p_1)+E(p_2),\qquad E(p) = 4 \sin^2(\frac{p}{2}) - 8 g^2 \sin^4(\frac{p}{2}) \\ S(p_2,p_1) = S_0 + g^2 S_2 \nonumber \\
S_0= -\frac{e^{ip_1+ip_2}-2e^{i p_2}+1}{e^{ip_1+ip_2}-2e^{i p_1}+1} \\
S_2= \frac{8 i e^{i p_1+i p_2} \sin(\frac{p_1}{2})( \sin (\frac{p_1- 3 p_2}{2})- 4 \sin (\frac{p_1-p_2}{2})+\sin (\frac{3 p_1-  p_2}{2}))\sin(\frac{p_2}{2})}{(1-2 e ^{i p_1}+ e ^ {ip_1 +i p_2})^2}. \end{gather} To two-loop order, this agrees with the solution given by equations (3.3) and (6.4), (4.27), and (3.7) of \cite{Staudacher:2004tk}. 

At this point, assuming diffractionless scattering and requiring periodicity yields the Bethe equation for this sector, which can be used to compute anomalous dimensions for states with arbitrary numbers of excitations, as in \cite{Staudacher:2004tk}.  As shown in Table \ref{tab:sl2}, we find perfect agreement between the predictions of the Bethe ansatz and direct diagonalization of the two-loop dilatation generator. This provides compelling evidence for two-loop integrability in the bosonic $\alsl(2)$ subsector.

\begin{table}\centering
$\begin{array}{|l|l|l|}\hline
D_0&(R, L, B) &\bigbrk{\algdD_2,\algdD_4}^P\\\hline
7 & (3, 3, 0)& \begin{array}{l} (6, -\frac{39}{4})^-  \end{array} \\ \hline
7 & (4, 4, 0)& \begin{array}{l} (6, -\frac{21}{2})^\pm \end{array} \\ \hline
8 & (3, 3, 0)& \begin{array}{l} (\frac{35}{4}, -\frac{18865}{1152})^\pm \end{array} \\ \hline
8 & (4, 4, 0)& \begin{array}{l} (4.38277, -5.25026)^+ \\ (8.35923, -16.0680)^+ \\ (11.5913, -23.1031)^+ \\ (\frac{23}{3}, -\frac{1331}{108})^\pm \end{array} \\ \hline
8 & (5, 5, 0)& \begin{array}{l} (4.72931,-7.01464)^\pm \\ (7.77069, -14.4229)^\pm \end{array} \\ \hline \end{array}$
\caption{Two-loop spectrum of highest weight states in the bosonic $\alsl(2)$ sector(s) found by direct diagonalization. }
\label{tab:sl2}
\end{table}

\subsection{The $\alSU(1|1)$ sector}
Finally, we provide evidence of integrability including fermions as well. We compute anomalous dimensions for the $\alSU(1|1)$ sector(s), again via direct diagonalization. This sector includes states made of only one type of $\phi$ and only one type of $\psi$, and no derivatives.  Again, our findings are in complete agreement with those found assuming integrabilty in \cite{Staudacher:2004tk}. These anomalous dimensions were also found by direct diagonalization of the compact $\alSU(2|3)$ dilatation operator in \cite{Beisert:2003ys}.

\begin{table}\centering
$\begin{array}{|l|l|l|}\hline
D_0&(R, L, B)) &\bigbrk{\algdD_2,\algdD_4}^P\\\hline
7  &(1,5,\pm 2)& \begin{array}{l} (10,-20)^- \end{array} \\
\hline 
7.5  &(3,6,\pm 1.5)& \begin{array}{l} (8,-14)^\pm \end{array} \\
\hline 
8  &(2,6,\pm 2)& \begin{array}{l} (8,-14)^+ \end{array} \\
\hline 
8.5  &(4,7,\pm 1.5)& \begin{array}{l} (7,-12)^\pm \end{array} \\
\hline 
9  &(3,7,\pm 2)& \begin{array}{l} (6.39612,-9.3993)^- \\ (9.10992, -17.1028)^- \\ (12.494, -24.4979)^- \end{array} \\
\hline 
9.5  &(5,8,\pm 1.5)& \begin{array}{l} (6,-\frac{19}{2})^\pm \\ (8, -\frac{29}{2})^\pm \end{array} \\
\hline \end{array}$
\caption{Two-loop spectrum of states in the $\alSU(1|1)$ sector found by direct diagonalization. }
\label{tab:anotab}
\end{table}

\section{Conclusion and outlook \label{sec:outlook}}
We have found a remarkably simple solution for the two-loop dilatation generator and the one and one-half loop symmetry algebra of a non-compact sector of $\mathcal{N}=4$ SYM. The $\mathcal{O}(g^2)$ and $\mathcal{O}(g^3)$ symmetry algebra corrections are given, with appropriate choice of basis and gauge, by 
\begin{equation} \algX^\pm_{i+2}=\pm  \comm{\algX^\pm_i}{\algx}, \quad i=0,\,1. \end{equation}
$\algx$, given by (\ref{eq:definex}), only involves the leading $\mathcal{O}(g^1)$ terms for the $\alPSU(1|1)^2$ generators and the harmonic numbers that characterize the one-loop dilatation operator. Furthermore, the two-loop dilatation generator (\ref{eq:D4solution}) generates the two-loop $\alsl(2)$ S-matrix of the Bethe ansatz proposal \cite{Staudacher:2004tk}, and its anomalous dimensions match both this Bethe ansatz proposal and the direct field theory calculations of \cite{Beisert:2003tq, Kotikov:2003fb, Eden:2005bt, Belitsky:2005bu}. This is additional very strong evidence in favor of the two-loop integrability and Bethe ansatz for the $\alSU(1,1|2)$ sector \cite{Beisert:2005fw}.

Despite the convincing evidence that our solution would be produced by a complete direct field theory computation, it is unknown if the solution is uniquely determined by symmetry and Feynman diagram constraints, as is the one-loop dilatation operator \cite{Beisert:2004ry}. It would be very interesting to identify the order at which the solution is not completely constrained (if any), and the minimal set of additional constraints required to isolate the field theory solution. 

The structure of the solution suggests additional directions of research. It is natural to conjecture that the iterative solution we found can be extended to larger sectors, which necessarily have dilatation generator interactions that do not conserve length and hypercharge, or to higher loops. Such a solution for the four-loop dilatation generator may be especially useful. At that order, non-trivial wrapping interactions for two-site states could be consistent with the symmetry algebra. Because the wrapping interactions are less constrained by inspecting Feynman diagrams, it would seem impossible to compute them just using the constraints. However, it is possible that a higher-loop extension of our iterative solution would not need specific wrapping interactions added to match the field theory solution. 

We suspect that there are expressions involving the same building blocks for the higher conserved charges due to integrability. Evidence for this was given in \cite{Agarwal:2005jj}. Up to two loops, Agarwal and Ferretti showed that the first higher charge for the $\alSU(2|3)$ sector could be written diagrammatically in terms of the dilatation generator. They conjectured that the diagrammatic expression generalizes to the entire theory. Using the solution for the two-loop dilatation generator, it is now possible to check whether their solution \cite{Agarwal:2005jj} generalizes to a non-compact sector.

Finally, this is not the first time iterative structures have appeared in $\mathcal{N}=4$ SYM.  Planar scattering amplitudes have iterative structure at two and three loops \cite{Anastasiou:2003kj, Bern:2005iz}. Also, following Witten's work relating gauge theory to a string theory in twistor space \cite{Witten:2003nn}, recurrence relations between amplitudes involving different numbers of particles have been found \cite{Cachazo:2004kj, Britto:2004ap}.  The two-loop dilatation generator has some qualitative resemblance to this recursive structure. The $\algT_1$ are analogous to the three gluon on-shell amplitudes, and $\algh$ is similar to a Feynman propogator. It would be wonderful if iterative structures could be used to relate the dilatation generator and scattering amplitudes.

\section*{Acknowledgements}
I thank my advisor, Niklas Beisert, for pointing me to the $\alSU(1,1|2)$ sector, for sharing essential $\texttt{Mathematica}$ code, for suggesting many improvements to this article and for numerous enlightening and stimulating conversations.  

I also thank Joshua Friess for helpful suggestions. This material is based upon work supported under a National Science Foundation Graduate Research Fellowship. Any opinions, findings, conclusions or recommendations expressed in this publication are those of the author and do
not necessarily reflect the views of the National Science Foundation.

\appendix
\section{The $\alPSU(1,1|2)$ commutators \label{sec:psu112comm}}

\subsection{Matrix representation}
The matrix representation  for the full $\alPSU(2,2|4)$ algebra shown in Appendix D of \cite{Beisert:2004ry} can be restricted to the $\alPSU(1,1|2)$ sector. 
As for the full algebra, we parameterize an element $j\cdot \algJ$ of the algebra 
by the adjoint vector $j$. In this case, we split the matrix into
$1|2|1$ (even$|$odd$|$even) rows and columns. We write the representation of $\alU(1,1|2)$ as

\[\label{eq:U112matrix2}
j\cdot \algJ=\left( \begin{array}{c|cc|c}
j^0 + b - \frac{c}{2} &\overrightarrow{q}^{+\downarrow}&\overrightarrow{q}^{+\uparrow}&j^{++}\\\hline 
\overleftarrow{q}^{-\uparrow}& r^0+\frac{b}{2} - \frac{c}{2}& r^{\uparrow\uparrow} &\overleftarrow{q}^{+\uparrow}\\ \overleftarrow{q}^{-\downarrow}&r^{\downarrow\downarrow}&  -r^0+\frac{b}{2} - \frac{c}{2}&\overleftarrow{q}^{+\downarrow}\\\hline
-j^{--}&-\overrightarrow{q}^{-\downarrow}&-\overrightarrow{q}^{-\uparrow}&-j^0+b - \frac{c}{2}  \end{array} \right)
\]
The commutation relations of the generators follow
from the matrix representation of $\comm{j\cdot \algJ}{j'\cdot \algJ}$. The $\alPSU(1,1|2)$ algebra follows from dropping $\algB$, which is not generated by any commutators of the other generators, and setting the central charge $\algC$ to zero. All physical fields are neutral with respect to $\algC$. 

We now present a minimal set of commutators; the remaining commutators follow from hermitian conjugation and from combining commutators presented here. We group these commutators for later convenience.

\subsection{Classical commutators}
These commutators only involve generators that receive no quantum corrections.
\begin{equation}  \comm{\algR^{\uparrow\uparrow}}{\algR^{\downarrow\downarrow}}=\algR^0, \qquad  \comm{\algR^0}{\algR^{\uparrow\uparrow}}=2 \algR^{\uparrow\uparrow}. \end{equation}

\subsection{Central charge commutators}
These commutators are simpler to deal with at higher orders in $g$, because $\algR^0$ receives no quantum corrections, and because \begin{equation} \algJ^0_n=\algdD_n \qquad n>0. \label{eq:J0nDn} \end{equation}
\begin{align} \comm{\algJ^0(g)}{\algJ^{++}(g)}&=2\algJ^{++}(g),& 
\comm{\algR^0}{\algJ^{++}(g)}&=0, \notag \\
\comm{\algJ^0(g)}{\overrightarrow{\algQ}^{+\downarrow}(g)}&= \overrightarrow{\algQ}^{+\downarrow}(g), &
\comm{\algR^0}{\overrightarrow{\algQ}^{+\downarrow}(g)}&=-\overrightarrow{\algQ}^{+\downarrow}(g), \notag \\
\comm{\algJ^0(g)}{\overleftarrow{\algQ}^{+\uparrow}(g)}&= \overleftarrow{\algQ}^{+\uparrow}(g), &
\comm{\algR^0}{\overleftarrow{\algQ}^{+\uparrow}(g)}&=\overleftarrow{\algQ}^{+\uparrow}(g) \notag \\
\comm{\algJ^0(g)}{\algR^{\uparrow\uparrow}}&=0.\end{align}

\subsection{Commutators with $\algR$}
\begin{align} 
\comm{\algR^{\uparrow\uparrow}}{\overleftarrow{\algQ}^{+\downarrow}(g)}&=\overleftarrow{\algQ}^{+\uparrow}(g) &
\comm{\algR^{\uparrow\uparrow}}{\overleftarrow{\algQ}^{-\downarrow}(g)}&=\overleftarrow{\algQ}^{-\uparrow}(g), \notag \\
\comm{\algR^{\uparrow\uparrow}}{\overrightarrow{\algQ}^{-\uparrow}(g)}&=0, &
\comm{\algR^{\uparrow\uparrow}}{\overrightarrow{\algQ}^{-\downarrow}(g)}&=-\overrightarrow{\algQ}^{-\uparrow}(g), \notag \\
\comm{\algR^{\uparrow\uparrow}}{\algJ^{++}(g)}&=0, &
\comm{\algR^{\uparrow\uparrow}}{\algJ^{--}(g)}&=0. \label{eq:R+comm} \end{align}

\subsection{Plus-plus commutators} 
\begin{align} 
\comm{\algJ^{++}(g)}{\overrightarrow{\algQ}^{+\downarrow}(g)}&=0,& 
\comm{\algJ^{++}(g)}{\overrightarrow{\algQ}^{+\uparrow}(g)}&=0, \notag \\
\comm{\algJ^{++}(g)}{\overleftarrow{\algQ}^{+\uparrow}(g)}&=0, &
\comm{\algJ^{++}(g)}{\overleftarrow{\algQ}^{+\downarrow}(g)}&=0, \notag \\
\acomm{\overrightarrow{\algQ}^{+\downarrow}(g)}{\overrightarrow{\algQ}^{+\downarrow}(g)}&=0, &  \acomm{\overrightarrow{\algQ}^{+\uparrow}(g)}{\overrightarrow{\algQ}^{+\uparrow}(g)}&=0, \notag \\
\acomm{\overleftarrow{\algQ}^{+\downarrow}(g)}{\overleftarrow{\algQ}^{+\downarrow}(g)}&=0, &  \acomm{\overleftarrow{\algQ}^{+\uparrow}(g)}{\overleftarrow{\algQ}^{+\uparrow}(g)}&=0. \end{align}
 
\subsection{$\algdD$ commutators}
These commutators, at non-zero order in $g$, yield a multiple of $\algdD$.
\begin{align} \comm{\algJ^{--}(g)}{\algJ^{++}(g)}&=\algJ^0(g),& 
\acomm{\overrightarrow{\algQ}^{+\downarrow}(g)}{\overleftarrow{\algQ}^{-\uparrow}(g)}&=\frac{1}{2}\algJ^0(g)+\frac{1}{2}\algR^0, \notag \\
\acomm{\overrightarrow{\algQ}^{+\uparrow}(g)}{\overleftarrow{\algQ}^{-\downarrow}(g)}&=\frac{1}{2}\algJ^0(g)-\frac{1}{2}\algR^0, & 
\acomm{\overleftarrow{\algQ}^{+\downarrow}(g)}{\overrightarrow{\algQ}^{-\uparrow}(g)}&=\frac{1}{2}\algJ^0(g)+\frac{1}{2}\algR^0, \notag \\
\acomm{\overleftarrow{\algQ}^{+\uparrow}(g)}{\overrightarrow{\algQ}^{-\downarrow}(g)}&=\frac{1}{2}\algJ^0(g)-\frac{1}{2}\algR^0.  \end{align} 

\subsection{Plus-minus commutators}
\begin{align}
\acomm{\overrightarrow{\algQ}^{+\downarrow}(g)}{\overleftarrow{\algQ}^{-\downarrow}(g)}&=\algR^{\downarrow\downarrow}, &
\acomm{\overrightarrow{\algQ}^{+\downarrow}(g)}{\overrightarrow{\algQ}^{-\uparrow}(g)}&=0, \notag \\
\comm{\algJ^{++}(g)}{\overleftarrow{\algQ}^{-\uparrow}(g)}&=-\overleftarrow{\algQ}^{+\uparrow}(g), &
\comm{\algJ^{++}(g)}{\overrightarrow{\algQ}^{-\downarrow}(g)}&=-\overrightarrow{\algQ}^{+\downarrow}(g). \label{eq:+-comm} \end{align}

\section{Proof of the $\mathcal{O}(g^2)$ solution \label{sec:g2proof}}
We will verify that (\ref{eq:g2solution}) is a solution by checking representatives of the minimal set of commutators given in the last section. We first present some necessary identities. Throughout this proof, we set $\algy$, the similarity transformation, to zero, without loss of generality. 

\subsection{Properties of the $\mathcal{O}(g^2)$ solution \label{sec:g2properties}}  Since $\algh$ commutes with $\algJ^0_0$ and the $\algR$'s, we have
\begin{align} \comm{\algx}{\algJ^0_0}&=0, & \comm{\algx}{\algR^{\uparrow\uparrow}}&=0, \notag \\
\comm{\algx}{\algR^{\downarrow\downarrow}}&=0, & \comm{\algx}{\algR^0}&=0. \label{eq:J00x} \end{align} Since $\algh$ commutes with $\len$, by (\ref{eq:psU11^2Alg}) \begin{equation} \comm{\algx}{\len}=0. \label{eq:xl} \end{equation} Next, we define
\begin{align} \comm{\algJ^{++}}{\algh}&=\algj^{++}, &  \comm{\overrightarrow{\algQ}^{+\downarrow}}{\algh}&=\overrightarrow{\algq}^{+\downarrow}, \notag \\
\comm{\overrightarrow{\algQ}^{+\uparrow}}{\algh}&=\overrightarrow{\algq}^{+\uparrow}, & \comm{\overleftarrow{\algQ}^{+\uparrow}}{\algh}&=\overleftarrow{\algq}^{+\uparrow}, \notag \\
\comm{\overleftarrow{\algQ}^{+\downarrow}}{\algh}&=\overleftarrow{\algq}^{+\downarrow} \label{eq:lowercase} \end{align}
As usual, the conjugates of these new generators are labeled with a minus instead of plus, and arrows pointing in the opposite direction. A useful simplification depends on the following vanishing commutators. 
\begin{align} \acomm{\overrightarrow{\algq}^{+\downarrow}}{\overrightarrow{\algT_1}^-}&=0, & \acomm{\overrightarrow{\algq}^{+\uparrow}}{\overrightarrow{\algT_1}^-}&=0, \notag \\
\acomm{\overleftarrow{\algq}^{+\uparrow}}{\overleftarrow{\algT_1}^-}&=0, & \acomm{\overleftarrow{\algq}^{+\downarrow}}{\overleftarrow{\algT_1}^-}&=0. \label{eq:vanishing} \end{align}
Then, from (\ref{eq:g2solution}) and (\ref{eq:definex}) we find
\begin{align} \overrightarrow{\algQ_2}^{+\downarrow}&=\comm{\overleftarrow{\algT_1}^-}{\acomm{\overrightarrow{\algT_1}^+}{\overrightarrow{\algq}^{+\downarrow}}}, & \overrightarrow{\algQ_2}^{+\uparrow}&=\comm{\overleftarrow{\algT_1}^-}{\acomm{\overrightarrow{\algT_1}^+}{\overrightarrow{\algq}^{+\uparrow}}}, \notag \\
\overleftarrow{\algQ_2}^{+\uparrow}&=\comm{\overrightarrow{\algT_1}^-}{\acomm{\overleftarrow{\algT_1}^+}{\overleftarrow{\algq}^{+\uparrow}}}, & \overleftarrow{\algQ_2}^{+\downarrow}&=\comm{\overrightarrow{\algT_1}^-}{\acomm{\overleftarrow{\algT_1}^+}{\overleftarrow{\algq}^{+\downarrow}}}. \label{eq:Q2} \end{align}

The following equalities will be essential.
\begin{align} \comm{\algJ^{++}_0}{\algj^{--}}&=\half \len, & \acomm{\overrightarrow{\algQ_0}^{+\downarrow}}{\overleftarrow{\algq}^{-\uparrow}}&=\frac{1}{4}(2 \algB - \len -\algR^0), \notag \\
\acomm{\overrightarrow{\algQ_0}^{+\uparrow}}{\overleftarrow{\algq}^{-\downarrow}}&=\frac{1}{4}(2 \algB - \len + \algR^0), &
\acomm{\overleftarrow{\algQ_0}^{+\downarrow}}{\overrightarrow{\algq}^{-\uparrow}}&=-\frac{1}{4}(2 \algB + \len +\algR^0), \notag \\
\acomm{\overleftarrow{\algQ_0}^{+\uparrow}}{\overrightarrow{\algq}^{-\downarrow}}&=-\frac{1}{4}(2 \algB + \len - \algR^0). \label{eq:essential} \end{align}

Finally, we have \begin{equation} \acomm{\overrightarrow{\algQ_0}^{+\downarrow}}{\overleftarrow{\algq}^{-\downarrow}}=-\half \algR^{\downarrow\downarrow}, \quad \comm{\algJ^{++}_0}{\overleftarrow{\algq}^{-\uparrow}}=0 \quad \mbox{and} \quad \comm{\algJ^{++}_0}{\overleftarrow{\algq}^{-\downarrow}}=0. \label{eq:finally} \end{equation}

\subsection{Central charge commutators}
We check the first commutator in detail.
\begin{eqnarray} \comm{\algJ^0}{\algJ^{++}}_2 &=& \comm{\algJ^0_0}{\algJ^{++}_2} + \comm{\algJ^0_2}{\algJ^{++}_0}\nonumber \\
&=& \comm{\algJ^0_0}{\comm{\algJ^{++}_0}{\algx}}+ \comm{\algdD_2}{\algJ^{++}_0} \nonumber \\
&=& \comm{\comm{\algJ^0_0}{\algJ^{++}_0}}{\algx} + \comm{\algJ^{++}_0}{\comm{\algJ^0_0}{\algx}} \nonumber \\
&=& 2\comm{\algJ^{++}_0}{\algx} \nonumber \\
&=& 2 \algJ^{++}_2. \end{eqnarray}

In the second line, we used (\ref{eq:J0nDn}), and in the fourth line we used (\ref{eq:J00x}).

\subsection{Commutators with $\algR$}
Here is one example of these proofs.
\begin{eqnarray} \comm{\algR^{\uparrow\uparrow}}{\overleftarrow{\algQ}^{+\downarrow}}_2 &=& \comm{\algR^{\uparrow\uparrow}_0}{\overleftarrow{\algQ_2}^{+\downarrow}} \nonumber \\
&=& \comm{\algR^{\uparrow\uparrow}_0}{\comm{\overleftarrow{\algQ_0}^{+\downarrow}}{\algx}} \nonumber \\
&=& \comm{\comm{\algR^{\uparrow\uparrow}_0}{\overleftarrow{\algQ_0}^{+\downarrow}}}{\algx} \nonumber \\
&=& \comm{\overleftarrow{\algQ_0}^{+\uparrow}}{\algx} \nonumber \\
&=& \overleftarrow{\algQ_2}^{+\uparrow}. \end{eqnarray} 
We used (\ref{eq:J00x}) again, as well as the leading order part of (\ref{eq:R+comm}).

\subsection{Plus-plus commutators} \label{sec:++}
The commutators of two plus generators vanish at this order since they vanish at leading order. For example:
\begin{eqnarray} \acomm{\overrightarrow{\algQ}^{+\downarrow}}{\overrightarrow{\algQ}^{+\uparrow}}_2 &=& \acomm{\overrightarrow{\algQ_2}^{+\downarrow}}{\overrightarrow{\algQ_0}^{+\uparrow}}+\acomm{\overrightarrow{\algQ_0}^{+\downarrow}}{\overrightarrow{\algQ_2}^{+\uparrow}} \nonumber \\
&=& \acomm{\comm{\overrightarrow{\algQ_0}^{+\downarrow}}{\algx}}{\overrightarrow{\algQ_0}^{+\uparrow}}+\acomm{\overrightarrow{\algQ_0}^{+\downarrow}}{\comm{\overrightarrow{\algQ_0}^{+\uparrow}}{\algx}} \nonumber \\ 
&=& \comm{\acomm{\overrightarrow{\algQ_0}^{+\downarrow}}{\overrightarrow{\algQ_0}^{+\uparrow}}}{\algx} \nonumber \\
&=& 0. \end{eqnarray}

\subsection{$\algdD$ commutators}
We proved the first commutator in Section \ref{sec:g2discussion}. For the other four commutators one must change signs appropriately and use (\ref{eq:Q2}) and (\ref{eq:essential}).  

\subsection{Plus-minus commutators}

The commutator of $\overrightarrow{\algQ}^{+\downarrow}$ and $\overleftarrow{\algQ}^{-\downarrow}$ vanishes at $\mathcal{O}(g^2)$ as required by (\ref{eq:+-comm}). In fact, both \begin{equation} \label{eq:Q+Q/} \acomm{\overrightarrow{\algQ_2}^{+\downarrow}}{ \overleftarrow{\algQ_0}^{-\downarrow}}\quad \mbox{and} \quad  \acomm{\overrightarrow{\algQ_0}^{+\downarrow}}{ \overleftarrow{\algQ_2}^{-\downarrow}} \end{equation} vanish. We will prove this for the first commutator. The proof for the second follows the same steps. First we need
\begin{eqnarray} \acomm{\overrightarrow{\algq}^{+\downarrow}}{\overleftarrow{\algQ_0}^{-\downarrow}} &=& \acomm{\comm{\overrightarrow{\algQ_0}^{+\downarrow}}{\algh}}{\overleftarrow{\algQ_0}^{-\downarrow}} \nonumber \\
&=& \comm{\algR^{\downarrow\downarrow}}{\algh}+\acomm{\overrightarrow{\algQ_0}^{+\downarrow}}{\overleftarrow{\algq}^{-\downarrow}} \nonumber \\
&=& \acomm{\overrightarrow{\algQ_0}^{+\downarrow}}{\overleftarrow{\algq}^{-\downarrow}} \nonumber \\
&=& -\half \algR^{\downarrow\downarrow}. \end{eqnarray}
We used (\ref{eq:finally}) for the last line.
Using (\ref{eq:vanishing}), we show the first commutator in (\ref{eq:Q+Q/}) vanishes.
\begin{eqnarray}  \acomm{\overrightarrow{\algQ_2}^{+\downarrow}}{\overleftarrow{\algQ_0}^{-\downarrow}} &=& \acomm{\comm{\overleftarrow{\algT_1}^-}{\acomm{\overrightarrow{\algT_1}^+}{\overrightarrow{\algq}^{+\downarrow}}}}{\overleftarrow{\algQ_0}^{-\downarrow}} \nonumber \\
&=&
\acomm{\overleftarrow{\algT_1}^-}{\comm{\overrightarrow{\algT_1}^+}{\acomm{\overrightarrow{\algq}^{+\downarrow}}{\overleftarrow{\algQ_0}^{-\downarrow}}}} \nonumber \\
&=&
\half \acomm{\overleftarrow{\algT_1}^-}{\comm{\overrightarrow{\algT_1}^+}{-\algR^{\downarrow\downarrow}}} \nonumber \\
&=& 0. \end{eqnarray}

Using (\ref{eq:vanishing}), \ref{eq:Q2}), and \begin{equation} \acomm{\overrightarrow{\algQ_0}^{+\downarrow}}{\overrightarrow{\algq}^{-\uparrow}}=\acomm{\overrightarrow{\algq}^{+\downarrow}}{\overrightarrow{\algQ_0}^{-\uparrow}}, \end{equation} 
one can show that anti-commutator of $\overrightarrow{\algQ}^{+\downarrow}$ and $\overrightarrow{\algQ}^{-\uparrow}$ vanishes.

As a preliminary step for verifying the next plus-minus commutator, we compute
\begin{eqnarray} \comm{\overleftarrow{\algQ_0}^{-\uparrow}}{\algj^{++}} &=& \comm{\overleftarrow{\algQ_0}^{-\uparrow}}{\comm{\algJ^{++}_0}{\algh}} \nonumber \\
&=& \comm{\overleftarrow{\algQ_0}^{+\uparrow}}{\algh}-\comm{\algJ^{++}_0}{\overleftarrow{\algq}^{-\uparrow}} \nonumber \\
&=& \overleftarrow{\algq}^{+\uparrow}-\comm{\algJ^{++}_0}{\overleftarrow{\algq}^{-\uparrow}} \nonumber \\
&=& \overleftarrow{\algq}^{+\uparrow}. \end{eqnarray}

The last line works because of (\ref{eq:finally}). (\ref{eq:finally}) also implies that $\overleftarrow{\algQ}^{-\uparrow}_2$ and $\algJ^{++}_0$ commute. Using this and (\ref{eq:vanishing}), we find 
\begin{eqnarray} \comm{\overleftarrow{\algQ}^{-\uparrow}}{\algJ^{++}}_2 &=& \comm{\overleftarrow{\algQ_2}^{-\uparrow}}{\algJ^{++}_0} + \comm{\overleftarrow{\algQ_0}^{-\uparrow}}{\algJ^{++}_2} \nonumber \\
&=& \comm{\overleftarrow{\algQ_0}^{-\uparrow}}{\acomm{\overrightarrow{\algT_1}^-}{\comm{\overleftarrow{\algT_1}^+}{\algj^{++}}}}-\comm{\overleftarrow{\algQ_0}^{-\uparrow}}{\acomm{\overrightarrow{\algT_1}^+}{\comm{\overleftarrow{\algT_1}^-}{\algj^{++}}}} \nonumber \\
&=& \comm{\overrightarrow{\algT_1}^-}{\acomm{\overleftarrow{\algT_1}^+}{\comm{\overleftarrow{\algQ_0}^{-\uparrow}}{\algj^{++}}}}-\comm{\overrightarrow{\algT_1}^+}{\acomm{\overleftarrow{\algT_1}^-}{\comm{\overleftarrow{\algQ_0}^{-\uparrow}}{\algj^{++}}}} \nonumber \\
&=& \comm{\overrightarrow{\algT_1}^-}{\acomm{\overleftarrow{\algT_1}^+}{\overleftarrow{\algq}^{+\uparrow}}} \nonumber \\
&=& \overleftarrow{\algQ_2}^{+\uparrow}. \end{eqnarray}

Equivalent steps show that \begin{equation} \comm{\overrightarrow{\algQ}^{-\downarrow}}{\algJ^{++}}_2 = \overrightarrow{\algQ_2}^{+\downarrow}. \end{equation} 

\subsection{Implied commutators}
Since the $\mathcal{O}(g^2)$ solution we presented (\ref{eq:g2solution}) is hermitian (for $\algy$ set to zero), like the leading order representation of the generators, it immediately follows that all the conjugate equations to those of Appendix \ref{sec:psu112comm} are satisfied at this order.  The algebra then implies that the remaining commutators are satisfied. This completes our verification of the symmetry constraints.

\section{Proof of the $\mathcal{O}(g^3)$ solution \label{sec:g3proof}}
We will first check that the solution (\ref{eq:g3solution}) commutes with the $\alPSU(1,1|2)$ generators, and then we will check that it satisfies the $\alPSU(1|1)^2$ algebra. Repeatedly, we will use the vanishing of the $\mathcal{O}(g^1)$ commutators without further comment. Again, we set $\algy$ to zero without loss of generality.  We also set $\alpha$ to zero without loss of generality, since a solution of the symmetry algebra remain a solution after a coupling constant transformation.
 
\subsection{Commutators with the $\alPSU(1,1|2)$ generators}
 
\subsubsection{Central charge and $\algR$ commutators}
To prove that the commutators of $\algT$ with the $\algR$'s or $\algJ^0$ vanish we need only that the $\algR$'s receive no quantum corrections, (\ref{eq:J00x}), (\ref{eq:J0nDn}), and 
\begin{equation} \comm{\algT_1}{\algdD_2} = 0. \end{equation}
This equation can be inferred from (\ref{eq:psU11^2Alg}) or from the fact that $\algdD$ commutes with all other generators. 

\subsubsection{Plus-plus commutators} \label{sec:cross++}
The proof for the commutators of $\overleftarrow{\algT}^+$ or $\overrightarrow{\algT}^+$ with any of the plus  $\alPSU(1,1|2)$ generators works in the same manner as in Appendix \ref{sec:++}, since commuting with $\algx$ generates the first quantum correction to all generators involved.

\subsubsection{Minus-plus commutators}
We will show how these commutators vanish using $\overleftarrow{\algT}^-$. With appropriate switching of $\alPSU(1,1|2)$ generators, the same proofs work for $\overrightarrow{\algT}^-$.
Using (\ref{eq:Q2}), we find
\begin{eqnarray} \acomm{\overleftarrow{\algT}^-}{\overrightarrow{\algQ}^{+\downarrow}}_3 &=& \acomm{\overleftarrow{\algT_1}^-}{\overrightarrow{\algQ_2}^{+\downarrow}} + \acomm{\overleftarrow{\algT_3}^-}{\overrightarrow{\algQ_0}^{+\downarrow}} \nonumber \\
&=& \acomm{\overleftarrow{\algT_1}^-}{\overrightarrow{\algQ_2}^{+\downarrow}} - \acomm{\comm{\overleftarrow{\algT_1}^-}{\algx}}{\overrightarrow{\algQ_0}^{+\downarrow}} \nonumber \\
&=& \acomm{\overleftarrow{\algT_1}^-}{\overrightarrow{\algQ_2}^{+\downarrow}} + \acomm{\overleftarrow{\algT_1}^-}{\comm{\overrightarrow{\algQ_0}^{+\downarrow}}{\algx}} \nonumber \\
&=& 2 \acomm{\overleftarrow{\algT_1}^-}{\overrightarrow{\algQ_2}^{+\downarrow}} \nonumber \\
&=& 2 \acomm{\overleftarrow{\algT_1}^-}{\comm{\overleftarrow{\algT_1}^-}{\acomm{\overrightarrow{\algT_1}^+}{\overrightarrow{\algq}^{+\downarrow}}}} \nonumber \\
&=& 0. \end{eqnarray}

The last equality follows from the general identity, 
\begin{equation} \acomm{Q}{\comm{Q}{R}}=0 \quad \mbox{if} \quad Q^2=0. \end{equation}

The same reasoning shows that the commutator of $\overleftarrow{\algT}^-$ with $\overrightarrow{\algQ}^{+\uparrow}$ vanishes. 

Since $\overleftarrow{\algT}^-$ commutes with $\overrightarrow{\algT}^-$ and $\overleftarrow{\algT}^+$, (\ref{eq:vanishing}) and (\ref{eq:Q2}) imply that \begin{equation} \acomm{\overleftarrow{\algT}^-}{\overleftarrow{\algQ}^{+\uparrow}}_3 =0 \quad \mbox{and} \acomm{\overleftarrow{\algT}^-}{\overleftarrow{\algQ}^{+\downarrow}}_3 =0. \end{equation}

The commutator of $\overleftarrow{\algT}^-$ with $\algJ^{++}$ actually is already fixed to zero at this order because $\algJ^{++}$ is generated by $\overrightarrow{\algQ}^{+\downarrow}$ and $\overleftarrow{\algQ}^{+\uparrow}$. We show this using the commutators given in Appendix \ref{sec:psu112comm},
\begin{eqnarray} \acomm{\overrightarrow{\algQ}^{+\downarrow}}{\overleftarrow{\algQ}^{+\uparrow}} &=& \acomm{\overrightarrow{\algQ}^{+\downarrow}}{\comm{\overleftarrow{\algQ}^{-\uparrow}}{\algJ^{++}}} \nonumber \\
&=& \comm{\acomm{\overrightarrow{\algQ}^{+\downarrow}}{\overleftarrow{\algQ}^{-\uparrow}}}{\algJ^{++}} \nonumber \\
&=& \comm{\half\algJ^0+\half\algR^0}{\algJ^{++}} \nonumber \\
&=& \algJ^{++}. \end{eqnarray}

\subsubsection{Implied commutators}
Using hermiticity and closure of the algebra, one can conclude that at $\mathcal{O}(g^3)$ all of the $\alPSU(1|1)^2$ generators commute with all of the $\alPSU(1,1|2)$ generators. 

\subsection{Commutators among the $\alPSU(1|1)^2$ generators}
Using hermiticity, it will be sufficient to check the following equations:

\begin{gather} \comm{\len}{\algT^+}_3= \algT^+_3, \label{comml} \\
\acomm{\overrightarrow{\algT}^+}{\overleftarrow{\algT}^-}_4=\acomm{\overleftarrow{\algT}^+}{\overrightarrow{\algT}^-}_4,  \label{eq:halfdD} \\  
\acomm{\overrightarrow{\algT}^+}{\overrightarrow{\algT}^-}_4=0, \quad \mbox{and} \quad 
\acomm{\overrightarrow{\algT}^+}{\overleftarrow{\algT}^+}_4=0,  \\   
(\overrightarrow{\algT}^+)^2_4 = (\overleftarrow{\algT}^+)^2_4=0.  
 \end{gather}

Both sides of (\ref{eq:halfdD}) are equal to $\half \algdD_4$ by (\ref{eq:psU11^2Alg}).

\subsubsection{Commutators with $\len$}
(\ref{comml}) follows from the corresponding leading order commutator and (\ref{eq:xl}).

\subsubsection{$\algdD$ commutators}
In fact, \begin{equation} \acomm{\overrightarrow{\algT_3}^+}{\overleftarrow{\algT_1}^-}=\acomm{\overleftarrow{\algT_3}^+}{\overrightarrow{\algT_1}^-}. \label{eq:U+-T+-} \end{equation} With the conjugate equation, this implies (\ref{eq:halfdD}). 
Here is the proof of (\ref{eq:U+-T+-}).
\begin{eqnarray} \acomm{\overrightarrow{\algT_3}^+}{\overleftarrow{\algT_1}^-}&=&\acomm{\comm{\overrightarrow{\algT_1}^+}{\algx}}{\overleftarrow{\algT_1}^-} \nonumber \\
&=& \acomm{\comm{\overrightarrow{\algT_1}^+}{\acomm{\overrightarrow{\algT_1}^-}{\comm{\overleftarrow{\algT_1}^+}{\algh}}}}{\overleftarrow{\algT_1}^-} \nonumber \\
&=& \acomm{\comm{\overrightarrow{\algT_1}^-}{\acomm{\overleftarrow{\algT_1}^+}{\comm{\overrightarrow{\algT_1}^+}{\algh}}}}{\overleftarrow{\algT_1}^-} \nonumber \\
&=& \acomm{\overrightarrow{\algT_1}^-}{\comm{\overleftarrow{\algT_1}^+}{\acomm{\overleftarrow{\algT_1}^-}{\comm{\overrightarrow{\algT_1}^+}{\algh}}}} \nonumber \\
&=& \acomm{\overrightarrow{\algT_1}^-}{\comm{\overleftarrow{\algT_1}^+}{\algx}} \nonumber \\
&=& \acomm{\overrightarrow{\algT_1}^-}{\overleftarrow{\algT_3}^+}. \end{eqnarray}
To obtain the second, and second to last lines we have used (\ref{eq:definex}) and applied the algebraic identity \begin{equation} \comm{Q}{\acomm{Q}{S}}=0 \quad \mbox{if} \quad Q^2=0. \end{equation} The proof also uses repeatedly the vanishing of the commutators of non-conjugate $\algT_1$'s.

\subsubsection{Non-conjugate $\algT$ commutators}
The same reasoning as in the previous section works to show that $\overrightarrow{\algT}^+$ and $\overrightarrow{\algT}^-$ commute.
\begin{eqnarray} \acomm{\overrightarrow{\algT}^+}{\overrightarrow{\algT}^-}_4&=& \acomm{\overrightarrow{\algT_3}^+}{\overrightarrow{\algT_1}^-} +  \acomm{\overrightarrow{\algT_1}^+}{\overrightarrow{\algT_3}^-} \nonumber \\
&=& \acomm{\comm{\overrightarrow{\algT_1}^+}{\algx}}{\overrightarrow{\algT_1}^-} -  \acomm{\overrightarrow{\algT_1}^+}{\comm{\overrightarrow{\algT_1}^-}{\algx}} \nonumber \\
&=& \acomm{\comm{\overrightarrow{\algT_1}^+}{\algx}}{\overrightarrow{\algT_1}^-} +  \acomm{\comm{\overrightarrow{\algT_1}^+}{\algx}}{\overrightarrow{\algT_1}^-} \nonumber \\
&=& 2 \acomm{\comm{\overrightarrow{\algT_1}^+}{\algx}}{\overrightarrow{\algT_1}^-} \nonumber \\
&=& \acomm{\comm{\overrightarrow{\algT_1}^+}{\acomm{\overrightarrow{\algT_1}^-}{\comm{\overleftarrow{\algT_1}^+}{\algh}}}}{\overrightarrow{\algT_1}^-} \nonumber \\
&=& \acomm{\comm{\overrightarrow{\algT_1}^-}{\acomm{\overleftarrow{\algT_1}^+}{\comm{\overrightarrow{\algT_1}^+}{\algh}}}}{\overrightarrow{\algT_1}^-} \nonumber \\
&=& 0. \end{eqnarray}

For the commutator of $\overrightarrow{\algT}^+$ with $\overleftarrow{\algT}^+$, see appendices \ref{sec:++} or \ref{sec:cross++}.

The squares of $\overrightarrow{\algT}^+$ and $\overleftarrow{\algT}^+$ vanish at $\mathcal{O}(g^4)$ for the same reasons that $\overrightarrow{\algT}^+$ commutes with $\overleftarrow{\algT}^+$ at this order.


\bibliographystyle{nb}

\end{document}